\documentstyle[11pt]{article}
\begin{document}
\centerline{\bf\large SINGULARITY IN POTENTIAL, PERTURBATION AND
VARIATIONAL METHODS}

\vspace{1cm}

\centerline{Yi-Bing Ding$^{1,2,5}$, Xue-Qian Li$^{1,3}$ and Peng-Nian
Shen
$^{1,4}$}

\vspace{1cm}

\centerline{1. CCAST (World Laboratory), P.O. Box 8730, Beijing 100080,
China}

\vspace{0.3cm}

\centerline{2. Physics Department, Graduate School of Academia Sinica,
Beijing, 100039, China}

\vspace{0.3cm}

\centerline{3. Department of Physics, Nankai University, Tianjin,
300071,
China}

\vspace{0.3cm}

\centerline{4. Institute of High Energy Physics, Academia Sinica,
Beijing,
100039, China}

\vspace{0.3cm}

\centerline{5. Department of Physics, University of  Milan , 20133
Milan, Italy}

\vspace{1cm}

\baselineskip 22pt

\begin{center}
\begin{minipage}{12cm}
\centerline{Abstract}

In this work, we carefully study the energy eigen-values and splitting
of heavy quarkonia as there exist $1/r^3$  and
$\delta^3(\vec r)$ singular terms in the potential
which make a direct numerical solution of the Schr\"{o}dinger equation
impossible. We compare the results obtained in terms of perturbation and
variational methods with various treatments.

\end{minipage}
\end{center}

\vspace{2cm}

It is well known that the interactions between heavy quarks (or heavy
quark-antiquark) can be reasonably described by a combination of the
Coulomb-type potential which is induced by the single-gluon exchange and
the confinement potential. That form corresponds to the zero-th order potential.

As the first order corrections, the Breit-Fermi
potential can be derived directly from the elastic scattering amplitude
of the quark and antiquark, where the non-relativistic approximation is
employed \cite{Lif}.

In the derivation,  a non-relativistic expansion with respect to
$(|\vec p|/M_Q)^n\;(n\geq 1)$ powers is carried out and usually, one
only keeps terms up to $|\vec p|^2/M_Q^2$ and ignore the higher orders.
A Fourier transformation would produce
$\delta^3(\vec r)$ and $1/r^3$ terms in the configuration space.
The extra potential term which is related to quark spins can
be written in a general form as  \cite{Luch}
\begin{eqnarray}
\label{gen}
V_{SD}(r) &=& ( {\vec L\cdot \vec s_1\over 2m_1^2}+{\vec L\cdot\vec
s_2\over
2m_2^2})({-1\over r}{dS\over dr}+{4\alpha_s\over 3r^3})+{4\alpha_s\over
3}\cdot
{1\over m_1m_2}{\vec L\cdot\vec S\over r^3} \nonumber \\
&+& {4\alpha_s\over 3}\cdot {2\over 3m_1m_2}\vec s_1\cdot \vec s_2\cdot
4\pi\delta(\vec r) \nonumber \\
&+& {4\alpha_s\over 3}\cdot {1\over m_1m_2}
[3(\vec s_1\cdot \hat r)(\vec s_2\cdot \hat r)-\vec s_1\cdot\vec s_2]
\cdot {1\over r^3},
\end{eqnarray}
where $\vec S\equiv \vec s_1+\vec s_2$, $\hat r$ is a unit spatial
vector,
$\vec L$ is the relative orbital angular momentum of the quark-antiquark
system.

As noted, even though the Coulomb potential $1/r$ is also singular at $r\to 0$,
it is benign for solving the second-order differential equation.
In fact, the Coulomb potential corresponds to the
zero-th order of the non-relativistic expansion of the one-gluon-exchange
induced potential, and
its singular behavior is fully compensated by the small measure at vicinity
of $r\rightarrow 0$. By contraries, the extra $\delta^3(\vec r)$ and $1/r^3$
come from higher order (at least the first order) of the expansion. They are
more singular and make the equation unsolvable. Here the
unsolvability means that when one tries to numerically solve the
Schr\"{o}dinger equation with the potential including $1/r^3$
and $\delta^3(\vec r)$ in
expression (\ref{gen}), he would confront infinity, namely the solution
blows up due to the singular behavior of the potential
at vicinity of $r\rightarrow 0$, i.e., no physical solution can be
obtained. Landau pointed
out that as the potential is more singular than $1/r^2$, the equation is
unsolvable \cite{Landau}, obviously $1/r^3$ and $\delta^3(\vec r)$ are
more singular than $1/r^2$.

Analysis tells us that such high order singularities are
not physical, but brought up by artificial truncation of the
non-relativistic expansion, namely one only keeps to $|\vec p|^2/M_Q^2$
(or even higher, but finite orders). 

Gupta et al. suggested to take an alternative expansion with respect to
$|\vec p|/E$ instead of $|\vec p|/M_Q$ \cite{Gupta}. Obviously the
divergence for $|\vec p|\rightarrow\infty $ (or $r\rightarrow 0$) is
avoided,
because $|\vec p|/E\rightarrow 1$ as $|\vec p|\rightarrow\infty$.
However,
$E$ is an operator itself, while it exists at denominator of the
expression,
the equation is not solvable either unless one takes some extra
approximations
\cite{Gupta}.

As discussed above, the singularity is artificial because of the
non-relativistic approximation, so that in principle, if one deals with
the problem in a reasonable scheme, for example, the relativistic B-S
equation, such singularity would not appear at all. However, the
potential form which is non-relativistic, is most useful and convenient.
Many good zero-th order results have been achieved \cite {Cornell}, we have all
reasons to believe that the non-relativistic
Schr\"{o}dinger equation properly describes the
physics. When the first order correction related to quark spins are
considered, which determines the fine-structure and hyperfine
structure of the quarkonium,
the singularity problem emerges, and it turns the whole equation unsolvable.
Therefore, as long as we are working with the non-relativistic
Schr\"{o}dinger system, we are forced to face this unphysical
singularity problem. Similar singularity problem was discussed in previous
literatures. For example, Frank, Land and Spector analyzed the singular
potential \cite{Frank}, however, they mainly concerned the scattering
states instead of the bound states. In this work, we only focus on the
bound state problems.
Our goal is to correctly solve the problem and  obtain  the physical
quantities, even though there is the singularity.

So now the question is  can we obtain the eigen energies and
wavefunctions
which are sufficiently close to physical reality? Of course, the
closeness or
validity degree can be tested by comparing the obtained
results about the splitting, for example, with experimental data.

The effect of the $\delta^3(\vec r)$ term on the s-wave has been
carefully studied by
some authors \cite{Luch}, so readers are recommended to the original
work for details.

In this work, we only concentrate ourselves at $1/r^3$ singular term which does
not affect the s-wave (see below).

In this paper, we will obtain the eigen energies in both perturbative and variational
methods and then compare
the results,  we also discuss some alternative ways to modify the
potential form for avoiding the singularity which
have been employed in literatures.
Our numerical results show that there exist certain regions for those
possible parameters in the approaches where all the results are consistent
to each other with sufficient accuracy.\\

\noindent (i). Perturbative method.

The zero-th order potential does not include the singular parts of
(\ref{gen}),
so that we can easily solve the corresponding
Schr\"{o}dinger equation and obtain the zero-th order eigen energies and
wavefunctions. Then we are going on studying the first order equation
which involves the singular terms and obtain corresponding energies in terms
of perturbative method.

Let us investigate the singular terms related to
$1/r^3$ in eq.(\ref{gen}), we find that
$\vec L\cdot \vec S={1\over 2}[(\vec L+\vec S)^2-\vec L^2-\vec
S^2]$,
so for the S-state ($l=0$), its eigenvalue is zero, moreover, the
tensor-coupling term in (\ref{gen}) is proportional to an operator
$$ V_{tensor}=
[3(\vec s_1\cdot \hat r)(\vec s_2\cdot \hat r)-(\vec s_1\cdot\vec s_2)],
$$
which is sandwiched between two states which are solutions of
the zero-th order equation. If both sides are S-states, then
$$\int d\Omega <Y_{00}|V_{tensor}|Y_{00}>=0,$$
where $\Omega$ is the solid angle. The lowest non-zero value comes
from P-state, $<l=1|V_{tensor}|l=1>\neq 0$, and the operator can also
result in a small mixing between S-state and D-state, i.e.
$<l=0|V_{tensor}|l=2>\neq 0$. We are not going to discuss the details of
the mixing in this work.

Therefore, the singular $1/r^3$ terms do not bother the S-states, but only
affect the higher orbital-angular momentum states ($l\geq 1$), this enables us
to  avoid singularities in practical calculations (see below).
With the first order correction the Schr\"{o}dinger equation
for $l=1, S=1$ and $J=2,1,0$ ($\vec S=\vec s_1+\vec s_2$) is recast as
\begin{equation}
\label{Sch}
[{-1\over 2\mu}{1\over r^2}{d\over dr}(r^2{d\over dr})+{2\over 2\mu r^2}
-{4\alpha_s\over 3r}+{3\over 2m^2}({4\alpha_s\over 3}){1\over r^3}
\left(\begin{array}{c} 1 \\ -1 \\ -2
\end{array} \right)]R(r)=ER(r),
\end{equation}
where we only keep the $\vec L\cdot \vec S$ term (dropped out the tensor
coupling for our purpose of this study) and take $m_1=m_2=m\equiv M_Q$,
the reduced mass $\mu=m/2$.

In this study, our purpose is to investigate the effects of the singular
potential, so that in the Schr\"{o}dinger equation we deliberately omit
the confinements piece, such $\kappa r$ which is well behaved at $r\rightarrow
0$ for convenience. This treatment does not change the results of the
study, but greatly simplifies the calculation.
\\

\noindent (ii) The practical calculation in terms of perturbation.

For heavy quarkonia $c\bar c$, $b\bar b$, the coefficients of the $1/r^3$
is much smaller than that of $1/r$ which has a milder behavior at
$r\rightarrow 0$, so one can employ the perturbation method to calculate the
splitting. Then the zero-th order equation is
\begin{equation}
({-1\over 2\mu}{1\over r^2}{d\over dr}(r^2{d\over dr})+{1\over \mu r^2}
-{4\alpha_s\over 3r})R^{(0)}(r)=E^{(0)}R^{(0)}(r).
\end{equation}

For the 1p state, there is an analytical solution as
\begin{eqnarray}
\label{wave}
R^{(0)}(r) &=& {r\over 2\sqrt 6 a^{5/2}}e^{-r/a} \\
E^{(0)} &=& {-1\over 2a}({4\alpha_s\over 3})\cdot {1\over 4},
\end{eqnarray}
and the Bohr radius is
\begin{equation}
a={1\over\mu({4\alpha_s\over 3})}.
\end{equation}
If we take the conventional parameters $m_c=1.84$ GeV, $\alpha_s=0.39$,
then
$$ a=2.09030\; {\rm GeV^{-1}}, \; E^{(0)}=-0.031096\; {\rm GeV}. $$
We can evaluate the contribution of the $1/r^3$ term by means of
perturbation. The perturbative hamiltonian is
\begin{equation}
\delta H={3\over 2m^2}({4\alpha_s\over 3}){1\over r^3}.
\end{equation}
Using the zero-th order wavefunction (\ref{wave}), we have
\begin{eqnarray}
\label{int}
I_P &=& <P|\delta H| P>\propto \int R_P^{(0)}(r){3\over
2m^2}({4\alpha_s\over
3}){1\over r^3}R_P^{(0)}(r)r^2dr \nonumber \\
&=& {\alpha_s\over 12 a^3 m^2}.
\end{eqnarray}

It is worth of noting,  $r$ exists at numerator of
the $R^{(0)}(r)$ expression in (\ref{wave}), so that the integration converges.

Numerically, we obtain
\begin{equation}
I_P=0.00105105\; {\rm GeV}.
\end{equation}
For $2^3P_2,\; 2^3P_1$ and  $2^3P_0$ states, the energy shifts are
$I_P,\;
-I_P$ and $-2I_P$ respectively. These values are only 3$\sim$ 6\% of
$E^{(0)}$, therefore the perturbation is legitimate and one can be
convinced  that the results are reliable.\\

\noindent (iii) Variational method.

The Schr\"{o}dinger equation is not solvable if there is a potential term
$-r^{-s}$ with $s>2$ \cite{Landau} or $\delta^3(\vec r)$,
it is because the corresponding eigen-energy tends to
$-\infty$ and average radius turns to 0. "Particle falls to the point
$r=0$". Let us focus on the $1/r^3$ singularity.
Definitely, the $-\infty$ solution is not physical, but a virtual solution
of the equation with the singular potential. The key point is how to
systematically get rid of the virtual unphysical solution and achieve
the physical one.

The hamiltonian is
\begin{equation}
\label{hp}
h_P=-{1\over 2\mu}{1\over r^2}{d\over dr}(r^2{d\over dr})+{1\over \mu
r^2}
-{4\alpha_s\over 3r}-{3\over 2m^2}({4\alpha_s\over 3}){1\over r^3},
\end{equation}
where we take $J=1$ as an example.
If we employ the P-state wavefunction (\ref{wave}) as a trial function,
the average value is
\begin{equation}
\label{hp1}
<h_P>={-\alpha_s\over 3a}+{1\over 8\mu a^2}-{\alpha_s\over 12 m^2 a^3},
\end{equation}
where $a$ is a variational parameter in the expression. Later we need to
differentiate the expression with respect to $a$ and obtain the minimum.

For a $c\bar c$ bound state, the relation between $<h_P>$ and $a$
shows that as $a\rightarrow 0$,
$<h_P>\rightarrow
-\infty$, there is no lower bound. The corresponding average radius is
\begin{equation}
<r>=5a,
\end{equation}
so that as $a\rightarrow 0$, $<r>\rightarrow 0$. It indicates that
if one took the minimal value $-\infty$ as the binding energy,
the radius turns out to be
infinitesimal, it cannot exist in principle. It is
exactly
the Landau's conclusion. We can also prove that the conclusion is
independent of the adopted form of the trial functions.  This result 
indicates that the singular behavior of the potential near $r\rightarrow 0$
is transferred to $<h_P>$ at $a\rightarrow 0$.

Strictly speaking, the equation is of no solution in this case, does it
mean that the Schr\"{o}dinger equation cannot describe our physics picture?
As pointed out in previous section, this singularity is caused by the ill
behavior of the non-relativistic approximation near $r=0$. One can conjecture
that the given Schr\"{o}dinger equation can still correctly describe the
physics even with the unphysical singular terms,
but we have to solve it in a proper way. Namely since the singularity
at $r\rightarrow 0$ is artificial and unphysical, the boundless energy
corresponding to $a=0$ does not reflect physical reality, i.e.
the infinite $<h_P>$ at $a=0$ is not a physical solution and must be eliminated.

From another angle, as $a\rightarrow 0$, the trial function (\ref{wave})
has a limit as
\begin{equation}
\lim_{a\rightarrow 0}{r\over a^{5/2}}e^{-r/a}=0.
\end{equation}
The trial function tends to zero, the solution becomes trivial and does not
make any sense at all.

In fact, if one differentiate $<h_p>$ with respect to $a$,
besides the minimal value $-\infty$ at the boundary
$a=0$, $<h_P>$ has a local minimum at $a_1$ and it is
\begin{equation}
a_1={12+\sqrt{144-192\alpha_s^2}\over
32\alpha_s}\stackrel{\alpha_s=0.39}{=}
1.97832\; {\rm GeV^{-1}}.
\end{equation}
The corresponding average energy value is
\begin{equation}
E_{var}=-0.0322362 \;{\rm GeV}.
\end{equation}

When we compare this value $E_{var}$ with the result obtained in terms of
perturbation
\begin{equation}
E_{per}=E^{(0)}-I_P=-0.0321470 \;{\rm GeV},
\end{equation}
the deviation between $E_{var}$ and $E_{per}$ is only about $10^{-3}$.

For a real meson, $<r>$ cannot be zero, so the variational parameter $a$
must be non-zero. With this constraint, the boundless energy which
corresponds to $a\rightarrow 0$, is ruled out and we can enjoy the
reasonable solution determined by the local minimum. Actually, this local
minimum should correspond to the real physical value.

Even though we know the negative infinite minimal value of $E$ is not
physical and should be abandoned, in practical calculations on computer, it is
still bothering, therefore some plausible ways which may eliminate
the singularity from the concerned problem would be welcome.\\

\noindent (iv) Variational method and a "smear" form factor.

One may introduce a form factor to "smear" the singularity
\cite{McClary}.
The most widely adopted form factor is the error function $erf(\lambda
r)$.
It turns the spatial part of the spin-orbit coupling to
\begin{equation}
V_P^{(SM)}={3\over 2m^2}{4\alpha_s\over 3}erf(\lambda r)[{1\over r^3}
-{2\over\sqrt{\pi}}{1\over r}].
\end{equation}
For any finite $\lambda-$value, $V^{(SM)}_P$ is no longer singular as
\begin{equation}
\lim_{r\rightarrow 0}V_P^{(SM)}={8\over 3m^2}{\alpha_s\lambda^3\over
\sqrt{\pi}},
\end{equation}
which is finite as long as $\lambda$ is finite.
In the approach, the hamiltonian turns into a form
\begin{equation}
h_P^{(SM)}={-1\over 2\mu}{1\over r^2}{d\over dr}(r^2{d\over dr})+{1\over
\mu r^2}-{4\alpha_s\over 3r}-V_P^{(SM)}.
\end{equation}
Because it has a nice behavior at $r\rightarrow 0$,  in $<h_p>$
there is no more the troublesome negative infinity at $a\rightarrow 0$, and
one can safely handle it
with the variational method. For $\lambda=5,10,15\; {\rm GeV}$, we
obtain the minimum
of $<h_P^{(SM)}>$ which is also the minimal value for the  whole range
of
$0<a<\infty$, and tabulate the results in table. 1.\\

\centerline{Table 1.}
\begin{center}
\vspace{0.3cm}
\begin{tabular}{|c|c|c|c|}
\hline
$\lambda\; ({\rm GeV})$ & 5 & 10 & 15\\
\hline
$a\; ({\rm GeV^{-1}})$ & 1.97967 & 1.97868 & 1.97849 \\
\hline
$E$ (GeV) & -0.0322276 & -0.322339 & -0.0322352\\
\hline
\end{tabular}
\end{center}

\vspace{0.5cm}

For $\lambda=15\; {\rm GeV}$, the result is exactly the same as
$E_{var}$ achieved in
last section. But as $\lambda>15\; {\rm GeV}$, $<h_P^{(SM)}>$ has a
boundless negative minimal value for $a\rightarrow 0$
again. It means that as $\lambda$ in the form factor
is greater than $15\; {\rm GeV}$, the function cannot smear out the
singularity behavior at $r\rightarrow 0$. The meaning of introducing the
"smear" form factor  is that once we introduce a "smear form factor
whose parameter resides within a certain range", we do not need to worry about
the boundless minimal value at all,
namely then the local minimum is the minimal value
and finite, so that we can safely use the variational method to obtain
energies. But beyond the parameter range, the boundless minimal
value appears again, then we need to look for the local minimum as
the real solution, exactly in analog to the treatment of previous
section.\\

\noindent (v) The variational method and the "regularization".

(a) Another approach to handle the singularity is the so-called
"regularization"
\cite{Fla}. In this approach, one can "regularize" ${1\over r^n}$
to ${1\over (r+c)^n}$ ($c>0$), which obviously makes the potential
converge
at $r\rightarrow 0$. Thus $V_P$ is replaced by
\begin{equation}
\label{reg}
V_P^{(r)}={3\over 2m^2}({4\alpha_s\over 3}){1\over (r+c)^3},
\end{equation}
and $V^{(r)}_P\rightarrow V_P$ as $c\rightarrow 0$. We tabulate the
results corresponding to various $c-$values in Table.2.\\

\centerline{Table 2}
\vspace{0.3cm}

\begin{center}
\begin{tabular}{|c|c|c|c|}
\hline
$c\; ({\rm GeV^{-1}})$ & 0.1 & 0.05 & 0.035 \\
\hline
$a\; ({\rm GeV^{-1}})$ & 1.99094 & 1.98525 & 1.98333 \\
\hline
$E\; ({\rm GeV})$ & -0.0321330 & -0.0321808 & -0.0321963 \\
\hline
\end{tabular}
\end{center}
\vspace{0.5cm}

It is noted that the smaller $c$ is, the closer to $E_{var}$ and
values in Table.1 the $E-$values obtained by the "regularization" scheme
are. But as $c<0.03\; {\rm GeV^{-1}}$, the singular behavior of the
potential shows up again,
namely then $<h_P^{(r)}>$ would have a boundless negative minimal value
at $a\rightarrow 0$.

In fact, in sections (iv) and (v),
two different approaches are introduced to remedy
the singular behavior of the potential at $r=0$. Besides the variational
parameter $a$, there is an extra parameter $\lambda$ or $c$ being
introduced.
We notice that if the
concerned parameter falls in a certain range, the negative infinite
minimal
value of $<h_P>$ which corresponds to an extreme variational parameter
(usually
$a\rightarrow 0$) does not exist at all and a local minimum would give
the physical solution. However, once the parameter is beyond the allowed
range,
such negative infinite $<h_P>$ would appear again, in this case, one has
to
enforce a constraint condition as in
section (iii) for the pure variational method,
i.e. prior rule out the negative unphysical infinity and keep the
local minimum as reasonable solution.

(b) As the simplest "regularization" scheme, one can replace $V_P^{(r)}$
in  eq.(\ref{reg}) by
\begin{equation}
\label{step}
V_P^{(r1)}=\theta(r-r_0){3\over 2m^2}{4\alpha_s\over 3}{1\over r^3}.
\end{equation}
Obviously, it is an equivalent way to eq.(\ref{reg}) to get rid of the
singularity at $r\rightarrow 0$ and $r_0$ can be seen as an arbitrarily
introduced "cut-off" in the configuration space or a corresponding
$\Lambda_0\sim 1/r_0$ in momentum space. For the zeroth order $1^3P_1$ state the
correction of $V_P^{(r1)}$ can be calculated and the numerical results are
shown in Table.3.\\

\centerline{table 3.}

\vspace{0.3cm}
\begin{center}
\begin{tabular}{|l|c|c|c|c|c|}
\hline
$r_0$ (GeV$^{-1}$) & 0.20 & 0.15 & 0.10 & 0.08 & 0.068 \\
\hline
$a$ (GeV$^{-1}$) & 1.97926 & 1.97886 & 1.97857 & 1.97848 & 1.97844 \\
\hline
$E$ (GeV) & $-0.0322303$ & $-0.0322328$ & $-0.0322347$ & $-0.0322352$ &
$-0.0322355$ \\
\hline
\end{tabular}
\end{center}
\vspace{0.3cm}

To make more sense, it would be interesting to re-calculate the energy
corrections owing to the potential $V_P^{r1}$ for the $1^3P_1$ state
in perturbation and compare
\begin{equation}
E^{r1}\equiv
<V_P^{r1}>=\int_{r_0}^{\infty}dr[{-3\over 2m^2}{4\alpha_s\over 3}{1\over r^3}
(R^{(0)})^2r^2],
\end{equation}
with
\begin{equation}
\delta\equiv
<V_P^{r1}>=\int_{0}^{r_0}dr[{-3\over 2m^2}{4\alpha_s\over 3}{1\over r^3}
(R^{(0)})^2r^2].
\end{equation}
Now let us tabulate the results in Table 4 as\\

\centerline{Table 4.}
\vspace{0.3cm}
\begin{center}
\begin{tabular}{|l|c|c|c|c|c|}
\hline
$r_0$ (GeV$^{-1}$) & 0.2 &  0.15 & 0.1 & 0.08 & 0.068 \\
\hline
$E^{r1}$ (GeV) & $-0.0010465$ & $-0.0010485$ & $-0.0010499$ & $-0.0010503$
& $-0.0010505$ \\
\hline
$\delta$ (GeV) & $-4.51\times 10^{-6}$ & $-2.58\times 10^{-6}$ &
$-1.17\times 10^{-6}$ & $-7.5\times 10^{-7}$ & $-5.4\times 10^{-7}$ \\
\hline
R & 0.0043 & 0.0024 & 0.0011 & 0.0007 & 0.0005\\
\hline
\end{tabular}
\end{center}
In the table, the ratio R is defined as
$$R={\delta\over E^{r1}}.$$

\vspace{0.4cm}

The integration over r from 0 to $r_0$ denotes the part we ignore when taking
the "regularization" (\ref{step}) with the step function $\theta(r-r_0)$.
The ratio R reflects the relative
errors, and from Table 4, we can see the errors
brought by the "regularization" are about a few thousandths at most. Therefore,
according to the experimental accuracy, the scheme is very satisfactory.
Indeed, our results restrict the probable error ranges for using
the "regularization" or "smear" scheme to remedy the singularity.\\

\noindent (vi) With the Gupta's approach.

In Gupta's approach \cite{Gupta}, where an approximation
$\vec s\equiv \vec p\;'+\vec p$ being small is taken,  $\vec p,\vec p\;'$
are the three-momenta of the two quarks. $V_P$ is replaced by
\begin{equation}
V_P^{g}={4\alpha_s\over 3}{3\over 2}{f_1(r)\over r},
\end{equation}
where
\begin{equation}
f_1(r)=[1-(1+2mr)e^{-2mr}]/m^2r^2,
\end{equation}
obviously
\begin{equation}
\lim_{r\rightarrow 0}f_1(r)=2.
\end{equation}
Thus this modified $V_P^{(g)}$ only possesses the singular behavior of
$1/r$,
so that would not bring up any difficulties for solving the
Schr\"{o}dinger
equation by variational method. With this approach, we employ the
same trial function (\ref{wave}) and variational parameter $a$ to obtain
energy as
\begin{equation}
a_{var}^{(g)}=1.98570\; {\rm GeV^{-1}},\;\;\;\;{\rm and}\;\;\;\;
E^{(g)}=-0.0321867 \;{\rm GeV}.
\end{equation}

Instead, if we use the perturbation method, we obtain
\begin{equation}
E_{pert}^{(g)}=-0.0321085 \;{\rm GeV}.
\end{equation}
This is consistent with the values given in Table.1 and 2, as the errors
are
negligibly small.\\

\noindent (vii) Conclusion and discussion.

Based on the above calculations,  we can draw our conclusion as

(a) As Landau pointed out that an ill-extrapolation of the
non-relativistic expansion brings up an
artificial singular $r^{-3}$ which makes the Schr\"{o}dinger equation
unsolvable. Here the "unsolvable" means that it is impossible to
obtain an exact solution of the equation, so that one is forced to look for a
reasonable way to find the physical solution from the given hamiltonian
with such singular terms. Our purpose is to seek for a solution which would describe
the concerned physics (eigen-energies and eigen-wavefunctions) to a satisfactory
degree. The success of the potential model for charmonia encourages this
approach and stimulates further studies.

(b) In particular, the variational method without any special treatments
for the singularity would result in an infinite negative energy which
corresponds to $<h_P>=-\infty,\; a=0, \; <r>=0$. Obviously it is not
physical,
because it manifests itself as an infinitesimal radius meson,
i.e. a point-like
particle which would cause singularity as well known. Moreover, as the
variational parameter turns to zero, the trial function (\ref{wave}) is
zero, so it becomes trivial and unphysical, so must be ruled out.

The local minimum
which corresponds to non-zero variational parameter and average radius
indeed
gives rise to reasonable solution and should be physical. Our results
show
 a satisfactory consistency with that by other approaches, as expected.

(c) The "smear" and "regularization" approaches all introduce new
parameters whose physical meaning is to decrease the measure near vicinity of
$r\rightarrow 0$ and/or simply eliminate
a small neighborhood of $r=0$. Such schemes ("smear" or
"regularization") with certain parameters partly compensate the artificial
truncation of the non-relativistic expansion.
Our results show that for certain parameter ranges,
the obtained results are consistent with the perturbation. For future
application, one needs to adopt reasonable values for the  parameters,
concretely, $\lambda=15\;{\rm GeV}$ for the "smear" approach and
$c=0.035\;{\rm GeV^{-1}}$  and $r_0=0.068$ for the
"regularization".

Beyond the parameter ranges, the singular behavior appears
again.

It may only concern the automatic calculation procedure on computer for
obtaining the energies by variational method. Without  carefully programming,
the computation may overflow once it gets to the artificial unphysical
$a\rightarrow 0$ and refuses to look for the physical local minimum.
Therefore when one writes his program, he has to be careful about the possible
negative infinity solution.

But if we take the suggested schemes with the concerned parameters residing
in the allowed regions, we do not need to worry about the unphysical infinity
at all while programming.

(d) The wavefunction problem is not solved yet. One can use the
perturbation
method to obtain the first order wavefunction, but it needs to sum over
contributions from all possible zeroth order states, so practical
applications are
restricted. With the variational method one can obtain wavefunctions
easily,
but as we studied in another work \cite{Ding}, it is not a simple job.
Even
without the troublesome $1/r^3$ terms, it is difficult to obtain
wavefunctions
with satisfactory accuracy. Corresponding to the well-known potential
forms,
the Cornell, logarithmic etc. we have found that it is easy to obtain
very
accurate energy levels, but the errors for wavefunctions, in particular,
for their values at origin are rather large
unless we carefully choose the forms of the trial functions and use the
multi-variational parameters. Moreover, the form of trial functions and
number
of variational parameters also depend on the form of potential. In
\cite{Ding}
we only succeeded to select proper trial functions and number of
parameters
for each well-known potential forms. However, that is
only for the zeroth order situation, so when we consider the first order
correction in the potential, the situation is even worsened.

So how to determine the wavefunctions, especially the values at origin,
with
existence of $1/r^3$ terms, is an open question and we will investigate
it
in our future works.\\

\vspace{0.5cm}

\noindent {\bf Acknowledgments}

This work is partially supported by the National Natural Science
Foundation
of China (NNSFC).

\end{document}